\documentclass[prb, aps, amsmath, amssymb, showkeys, showpacs,twocolumn] {revtex4-1}
\usepackage[caption=false]{subfig}
\usepackage[colorlinks=false]{hyperref}
\usepackage{graphicx}
\usepackage{dcolumn}
\usepackage{bm}
\begin{document}
\title{Principle of detailed balance and a dilute gas in gravitational field}
\author{Kai Zhang}
\author{Yong-Jun Zhang}
\email{yong.j.zhang@gmail.com}
\affiliation{Science College, Liaoning Technical University, Fuxin, Liaoning 123000, China}

\begin{abstract}
We study in this paper a dilute gas in a gravitational field and present a relation of the molecular distribution function with respect to position and velocity. The relation is obtained from the principle of detailed balance and can be used to study temperature distribution and density distribution.
\end{abstract}
\keywords{detailed balance principle; dilute gas; gravity; temperature;}
\pacs{51.90.+r, 05.70.Ce, 02.50.Cw, 05.60.Cd}
\maketitle

\section{introduction}
Gibbs \cite{Gibbs}, using entropy method, proved that the temperature distribution of an object in an external gravitational field is uniform. The Boltzmann equation produces the same result \cite{BoltzmannEquation} for a dilute gas. But our intuition tells us that a molecule moving up in a gravitational field loses energy and therefore should lead to a lower temperature. With the similar reasoning, Josef Loschmidt claimed that {\it the equilibrium temperature of a gas column subject to gravity should be lower at the top of the column and higher at its base.} This is known as Loschmidt's gravito-thermal effect \cite{Trupp,book}, about which some experiments \cite{Graeff, Liao} have been carried out. However, as Maxwell noted, this effect violates the second law of thermodynamics. Then a paradox was raised by Sheehan {\it et al.} \cite{Sheehan} saying that gravitational field leads to effect violating the second law of thermodynamics. Wheeler \cite{Wheeler} examined the paradox and found that there is no violation.

Our intuition works better on kinetic theory instead of entropy theory. Therefore Coombes and Laue \cite{cccc} derived a relation about molecular distribution function with respect to height and velocity, with which, they were able to explain why the temperature remains uniform intuitively. However, their relation was obtained without taking into account the molecular collisions.

To take into account molecular collisions, we present a new derivation. We do this in two steps. In step one, we use the principle of detailed balance \cite{Tolman} to study only those processes involving no collisions. In step two, we use the principle of microscopic reversibility \cite{Lewis} and a thought experiment to study collisions. 

\section{detailed balance principle with no collisions}
\begin{figure*}
\subfloat[
A molecule at position $\vec{r}_1$ with velocity $\vec{v}_1$. When no collisions happen on the path, it reaches $\vec{r}_2$ with velocity $\vec{v}_2$\label{sfig:testa}]{%
	\includegraphics[width=8.2cm]{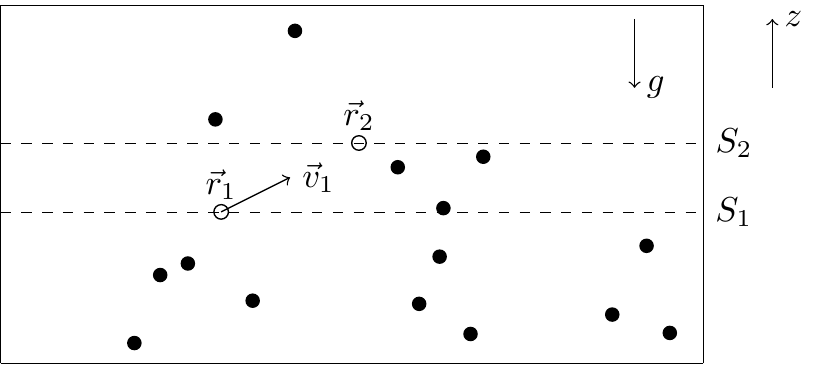}
}
\hfill
\subfloat[
A molecule at position $\vec{r}_2$ with velocity $-\vec{v}_2$. When no collisions happen on the path, it reaches $\vec{r}_1$ with velocity $-\vec{v}_1$\label{sfig:testa}]{%
	\includegraphics[width=8.2cm]{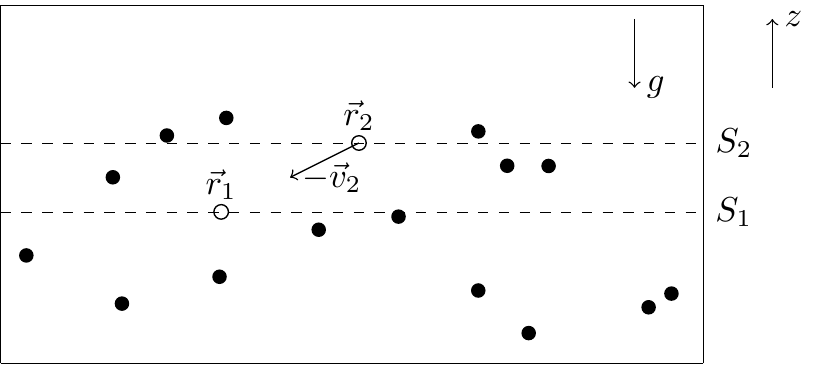}
}
\caption{A dilute gas in a gravitational field. 
Two cross sections $S_1$ and $S_2$ are also shown.
}
\label{gas}
\end{figure*}

Figure~\ref{gas} shows a dilute gas in equilibrium in a gravitational field with a constant acceleration $g$. Let us study a molecule at position $\vec{r}_1(x_1,y_1,z_1)$ with velocity $\vec{v}_1(v_{1x},v_{1y},v_{1z})$. In case of no collisions, the molecule reaches position $\vec{r}_2(x_2,y_2,z_2)$ with corresponding velocity $\vec{v}_2(v_{2x},v_{2y},v_{2z})$ where 
\begin{equation}\label{first} 
	\begin{array}{l}
	v_{2x}=v_{1x},\\
	v_{2y}=v_{1y},\\
	\frac{1}{2}mv_{2z}^2+mgz_2=\frac{1}{2}mv_{1z}^2+mgz_1,
	\end{array}
\end{equation}
and $m$ is the mass of the molecule.

Let us introduce a distribution function $f(\vec{r},\vec{v})$ that is defined in such a way that $f(\vec{r},\vec{v})d^3\vec{r}d^3\vec{v}$ is the number of molecules within phase space volume element $d^3\vec{r}d^3\vec{v}$ in the neighbourhood of ($\vec{r},\vec{v}$). The system being in equilibrium means that the distribution function $f(\vec{r},\vec{v})$ must obey some relation which we will find in two steps.

In step one, we study only those processes involving no collisions for a molecule moving from $\vec{r}_1$ to $\vec{r}_2$. For that, we write the distribution function into two parts,
\begin{eqnarray}\label{f12} 
	f(\vec{r}_1,\vec{v}_1)= &&f(\vec{r}_1,\vec{v}_1, \mathop{\vec{r}_1\longrightarrow \vec{r}_2}_{\text{collisions}}^{\text{no}})\nonumber\\
 +&& f(\vec{r}_1,\vec{v}_1, \mathop{\vec{r}_1\longrightarrow \cdots }_{\text{collisions}}^{\text{with}}).
\end{eqnarray}
When no collisions happen on the path, a molecule from $\vec{r}_1$ with $\vec{v}_1$ reaches $\vec{r}_2$ with $\vec{v}_2$. When collisions happen before it reaches $\vec{r}_2$, the molecule will deviate from the path and go somewhere else. In the same way, we write
\begin{eqnarray}\label{f21}
	f(\vec{r}_2,-\vec{v}_2)=&& f(\vec{r}_2,-\vec{v}_2, \mathop{\vec{r}_2\longrightarrow \vec{r}_1}_{\text{collisions}}^{\text{no}})\nonumber\\
 +&& f(\vec{r}_2,-\vec{v}_2, \mathop{\vec{r}_2\longrightarrow \cdots }_{\text{collisions}}^{\text{with}}).
\end{eqnarray}
In order to use the principle of detailed balance, imagine two cross sections $S_1$ and $S_2$ at heights $z_1$ and $z_2$, respectively (Fig.~\ref{gas}). 
Then we can introduce two quantities
\begin{eqnarray}\label{1} 
	&&\Delta F(\vec{v}_1, \mathop{S_1\longrightarrow S_2}_{\text{collisions}}^{\text{no}})\nonumber\\ 
=&& \iint\limits_{S_1} f(\vec{r}_1,\vec{v}_1, \mathop{\vec{r}_1\longrightarrow S_2}_{\text{collisions}}^{\text{no}})  \Delta v_{1x}\Delta v_{1y}\Delta v_{1z} dx_1dy_1
\end{eqnarray}
and
\begin{eqnarray}\label{2} 
&&	\Delta F(-\vec{v}_2, \mathop{S_2\longrightarrow S_1}_{\text{collisions}}^{\text{no}} )\nonumber \\ 
=&& \iint\limits_{S_2} f(\vec{r}_2,-\vec{v}_2, \mathop{\vec{r}_2\longrightarrow S_1}_{\text{collisions}}^{\text{no}} )\Delta v_{2x}\Delta v_{2y}\Delta v_{2z} dx_2dy_2,
\end{eqnarray}
where
\begin{equation}\label{3} 
	\Delta v_{2x}=\Delta v_{1x},\ \ \Delta v_{2y}=\Delta v_{1y},\ \ \Delta v_{2z}=\frac{v_{1z}}{v_{2z}}\Delta v_{1z},
\end{equation}
which is from Eq.~(\ref{first}). 

One can introduce a flux density \cite{Jeans, Chapman, Wheeler} $j(\vec{r},\vec{v})=v_z f(\vec{r},\vec{v})$ to construct a continuity equation. But here, given that the system is in equilibrium, we can use the principle of detailed balance \cite{Tolman} to write
\begin{equation}\label{4} 
	 {v}_{1z}\Delta F(\vec{v}_1, \mathop{S_1\longrightarrow S_2}_{\text{collisions}}^{\text{no}})= v_{2z}\Delta F(-\vec{v}_2, \mathop{S_2\longrightarrow S_1}_{\text{collisions}}^{\text{no}}),
\end{equation}
which means that in unit time, the average number of those molecules crossing $S_1$ with velocity $\vec{v}_1$ reaching $S_2$ with velocity $\vec{v}_2$ equals the average number of those molecules crossing $S_2$ with velocity $-\vec{v}_2$ reaching $S_1$ with velocity $-\vec{v}_1$. 

Combining all the equations together, we get
\begin{eqnarray}\label{int} 
	&&\iint\limits_{S_1}f(\vec{r}_1,\vec{v}_1, \mathop{\vec{r}_1\longrightarrow S_2}_{\text{collisions}}^{\text{no}})   dx_1dy_1 \nonumber\\
	=&&\iint\limits_{S_2}f(\vec{r}_2,-\vec{v}_2, \mathop{\vec{r}_2\longrightarrow S_1}_{\text{collisions}}^{\text{no}} )dx_2dy_2 .
\end{eqnarray}
In fact, the function $f(\vec{r},\vec{v})$ is uniform on $S_1$ and $S_2$ separately, so Eq. (\ref{int}) means
\begin{equation} \label{5}
	f(\vec{r}_1,\vec{v}_1, \mathop{\vec{r}_1\longrightarrow \vec{r}_2}_{\text{collisions}}^{\text{no}}) =f(\vec{r}_2,-\vec{v}_2,\mathop{\vec{r}_2\longrightarrow \vec{r}_1}_{\text{collisions}}^{\text{no}}).
\end{equation}
So far we have only considered those processes involving no collisions. We will show in the next section that a similar relation holds for all processes.

\section{Microscopic reversibility and collisions}
Now we study those processes involving collisions. We need to use a thought experiment. First, we look at the dilute gas in Fig.~\ref{gas} and observe how the molecules move and collide as time goes on. Then, at a random time, we put a dark matter \cite{darkmatter} molecule at $\vec{r}_1$ with $\vec{v}_1$. Since the dark matter molecule does not collide with any other molecules, it would always reach $\vec{r}_2$ with velocity $\vec{v}_2$. We do the thought experiment $N$ times. Suppose $N_1$ times the dark matter molecule passes through at least one ordinary molecule on the path.

In the same way, we do thought experiment of a dark matter molecule going from $\vec{r}_2$ with $-\vec{v}_2$ to $\vec{r}_1$. Again, we do the experiment $N$ times, and $N_2$ times the dark matter molecule passes through at least one ordinary molecule on the path.

In the limit of $N\to \infty$, we have 
\begin{equation} \label{NN}
	\lim_{N\to \infty}\frac{N_1}{N}= \lim_{N\to \infty}\frac{N_2}{N}.
\end{equation}
This is guaranteed by the principle of microscopic reversibility \cite{Lewis}: {\it Corresponding to every individual process there is a reverse process, and in a state of equilibrium the average rate of every process equals the average rate of its reverse process}. 
 In the limit of $N\to \infty$, for every microscopic process counted in $N_1$, there is a time reversal microscopic process counted in $N_2$, and vice versa. 

Now, we replace the dark matter molecule by an ordinary molecule and replicate each thought experiment. Whenever the dark matter molecule passes through the first molecule on its path, the ordinary molecule would collide with it and deviate from the path. Therefore in $N$ experiments about the ordinary molecule going from $\vec{r}_1$ with $\vec{v}_1$ to $\vec{r}_2$, $N_1$ of them would have collisions. So we know that the probability for a molecule to collide with other molecules on its path going from $\vec{r}_1$ with $\vec{v}_1$ to $\vec{r}_2$ is
\begin{equation} 
	P_1=\lim_{N\to \infty}\frac{N_1}{N}.
\end{equation}
In the same way, the probability for a molecule to collide with other molecules on its path going from $\vec{r}_2$ with $-\vec{v}_2$ to $\vec{r}_1$ is 
\begin{equation} 
P_2=\lim_{N\to \infty}\frac{N_2}{N}.
\end{equation}
Combining with Eq.~(\ref{NN}), we know that 
\begin{equation}\label{psymmetry} 
	P_1=P_2.
\end{equation}

From Eq.~(\ref{f12}) and Eq.~(\ref{f21}), we can write
\begin{equation} \label{e1}
	f(\vec{r}_1,\vec{v}_1, \mathop{\vec{r}_1\longrightarrow \vec{r}_2}_{\text{collisions}}^{\text{no}}) = f(\vec{r}_1,\vec{v}_1)(1-P_1),
\end{equation}
and
\begin{equation}\label{e2} 
	f(\vec{r}_2,-\vec{v}_2, \mathop{\vec{r}_2\longrightarrow \vec{r}_1}_{\text{collisions}}^{\text{no}}) = f(\vec{r}_2,-\vec{v}_2)(1-P_2).
\end{equation}
Substituting them into Eq.~(\ref{5}) and considering Eq.~(\ref{psymmetry}), we get
\begin{equation}\label{new_relation} 
	f(\vec{r}_1,\vec{v}_1)=f(\vec{r}_2,-\vec{v}_2),
\end{equation}
where $\vec{r}_1$, $\vec{r}_2$, $\vec{v}_1$ and $\vec{v}_2$ are related to each other as shown in Fig.~\ref{gas} and Eq.~(\ref{first}). This relation can be interpreted as a specific version of the principle of detailed balance for a dilute gas in a gravitational field with collisions taken into account.

\section{temperature and density}
The relation (\ref{new_relation}) determines the temperature distribution and the density distribution.  
In one-dimensional situation, it becomes 
\begin{equation} \label{zz}
	f(z_1,v_{1z})=f(z_2,-v_{2z}).
\end{equation}
If the Maxwell distribution holds locally, the relation further becomes $f(z_1,v_{1z})=f(z_2,v_{2z})$, which is the result that Coombes and Laue  \cite{cccc} have obtained in a different way without taking into account the molecular collisions. 
With their relation, Coombes and Laue were able to explain intuitively why the temperature is uniform and why the barometric formula \cite{Berberan-Santos} holds. 

We can do the same. If the Maxwell distribution holds, we can write
\begin{equation}\label{ff} 
	\begin{array}{l}
	f(z_1,v_{1z})=n(z_1)\sqrt{\frac{m}{2\pi kT_1}}\exp\left(-\frac{mv_{1z}^2}{2kT_1}\right),\\
	f(z_2,-v_{2z})=n(z_2)\sqrt{\frac{m}{2\pi kT_2}}\exp\left(-\frac{mv_{2z}^2}{2kT_2}\right),\\
	\frac{1}{2}mv_{2z}^2+mgz_2=\frac{1}{2}mv_{1z}^2+mgz_1,
	\end{array}
\end{equation}
where $n$ is the molecular number density and $T_1$ and $T_2$ are the temperatures at heights $z_1$ and $z_2$, respectively. Substituting them into Eq.~(\ref{zz}), we get
\begin{eqnarray} 
	&&\frac{n(z_1)}{n(z_2)}\exp\left[-\frac{mg(z_2-z_1)}{kT_2}\right] \nonumber\\
 	&&\times\sqrt{\frac{T_2}{T_1}}\exp\left[-\frac{mv_{1z}^2}{2k}\left(\frac{1}{T_1}-\frac{1}{T_2}\right)\right]=1.
\end{eqnarray}
This relation must hold for all $v_{1z}$. Here $z$ and $v_z$ are independent variables. So we have $T_1=T_2$. This means that the temperature $T$ is uniform and the barometric formula holds,
\begin{equation} 
 	\frac{n(z_2)}{n(z_1)}=\exp\left[-\frac{mg(z_2-z_1)}{kT}\right].
\end{equation}

According to Maxwell distribution, the temperature is determined by the shape of the velocity distribution function. A molecule going up against the gravitational field indeed loses energy. But this does not change the shape of velocity distribution function. So the temperature does not change. 
About the Maxwell distribution, it holds locally for an equilibrium state whether or not a gravitational field presents \cite{cccc}. But note that Rom\'{a}n, White and Velasco \cite{bbbb,aaaa} argued that the Maxwell distribution should hold only in the thermodynamic limit.

\begin{acknowledgments}
The authors are grateful to Shu-Hang Li and Jie Ren for discussions.
\end{acknowledgments}

\end{document}